\documentclass[a4paper,12pt]{article}
\usepackage{graphicx}

\usepackage{amsmath,amssymb,graphicx}
\usepackage{cite}
\usepackage{color}

\usepackage{color}
\usepackage{ifpdf}
\ifpdf 
  \DeclareGraphicsExtensions{.pdf,.png,.jpg,.jpeg,.mps}
  \usepackage{pgf}
  \usepackage{tikz}
\else 
  \usepackage{graphicx}
  \DeclareGraphicsExtensions{.eps,.bmp}
  \usepackage{pgf}
  \usepackage{tikz}
  \usepackage{pstricks}
\fi
\usepackage{epic,bez123}
\usepackage{floatflt}
\usepackage{wrapfig}

\numberwithin{equation}{section}

\newcommand{\ii}{\mathrm{i}}

\newcommand{\dd}{\mathrm{d}}
\newcommand{\pd}{\partial}

\newcommand{\e}{\mathrm{e}}

\newcommand{\diag}{\mathop{\mathrm{diag}}}

\newcommand{\I}{\mathbb{I}}

\newcommand{\ft}[2]{{\textstyle\frac{#1}{#2}}}

\begin{document}

\title{On the defect induced gauge and Yukawa fields in graphene.}%
\author{Corneliu Sochichiu\thanks{e-mail:
\texttt{sochichi@skku.edu}}\\
{\it University College,}\\
{\it Sungkyunkwan University, Suwon 440-746, KOREA}
\and
{\it Institutul de
    Fizic\u a Aplicat\u a A\c S,}\\
{\it  str. Academiei, nr. 5,
    Chi\c{s}in\u{a}u, MD2028, MOLDOVA}
}%
%

\maketitle
\begin{abstract}
We consider lattice deformations (both continuous and topological) in the hexagonal lattice Hubbard model in the tight binding approximation to graphene, involving operators with the range up to next-to-neighbor. In the low energy limit, we find that these deformations give rise to couplings of the electronic Dirac field to an external scalar (Yukawa) and gauge fields. The fields are expressed in terms of original defects. As a by-product we establish that the next-to-nearest order is the minimal range of deformations which produces the complete gauge and scalar fields. We consider an example of Stone--Wales defect, and find the associated gauge field.
\end{abstract}
\section{Introduction}
Graphene attracted a lot of interest in recent years from both theoretical and experimental communities due to its remarkable properties \cite{2009RvMP...81..109C}. Especially intriguing feature of graphene's electronic structure is that in the low energy limit it is described by a (pseudo)relativistic massless Dirac fermion model \cite{PhysRev.71.622}. This property provides a link between the condensed matter theory and high energy physics. Another interesting feature of graphene is its chirality. The microscopic theory is non-relativistic and chiral non-invariant, but due to a phenomenon similar to fermion doubling in lattice field theory, the chiral and Lorentz symmetries, as well as an internal symmetry emerge in the low energy (continuum) limit \cite{Semenoff:1984PhRvL}. 

The nonabelian character of the internal symmetry of the system is inspiring for a further quest into other links which graphene and high energy physics may have. In particular, an attractive idea is to discover the non-abelian gauge fields in graphene. A natural candidate for this role is provided by the phonon field  \cite{Kleinert1989b}, or more generally by the defects of the hexagon lattice.  Thus, experimentally it was discovered that the phonon couples to the electron wave function in an anomalous way. This lead to the suggestion, that this coupling should be realized, in fact, as a Dirac fermion/gauge or gravity coupling \cite{Pisana2007,CastroNeto2007,JPSJ.75.124701,PhysRevB.75.045404}. The phonon field can be regarded as a particular case of smooth defect. In more general case including topological defects, it was found that various types of lattice defects can be described in the low energy limit by coupling the fermionic field to various non-abelian field backgrounds \cite{Katsnelson20073,PhysRevB.65.235412,2005PThPh.113..463S} (see \cite{Vozmediano2010} for a review).

The objective of this work is to consider most arbitrary deformation of the hexagonal Hubbard model, corresponding to the inclusion of various local or short range defects.  The general problem related, in particular, to topological defects is that in many cases they do not admit, even in the low energy limit, a description in terms of continuous fields. In contrast, they may localize around several points of the momentum space. Our strategy in this work is to consider a generic setup and extract the modes coupled to the low energy electronic field. It appears, that it is the modes localized around the origin as well as around the Dirac points which are `seen' by the low energy fermionic field. 

The plan of the remainder of the paper is as follows. In the next section we give a brief introduction to the low energy theory of the free electronic wave function and introduce the notations. In the following section we consider the deformation of the hexagonal lattice Hubbard model by inserting the terms into the Hamiltonian corresponding to \emph{local}, \emph{nearest neighbor} and \emph{next-to-nearest neighbor} deformations of the model. We assume, that the deformation has an \emph{a priori} arbitrary spectral composition. Then we take the low energy limit and select the modes which couple to the low energy fermionic wave function. The main result is summarized by eq.~\eqref{final-coupling}: the defects emerge as a U(2) scalar (Yukawa) and U(2) gauge couping. The result also shows that the next-to-neighbor term is important to obtain the complete gauge/scalar coupling. At the same time this type of deformations is also saturating: adding more distant interaction terms to the Hamiltonian will not change the character of interactions in the low energy theory, but just will renormalize it. We find these properties of the model remarkable. We also rule out the assumption existent in the literature that the coupling can be fermionic gravity type coupling. In the Section  \ref{sec:SWD} we consider an application of the general theory to the description of the Stone--Wales defect in terms of gauge fields. 
 In addition we include two sections in the Appendix: One providing the complete information on the spinor and gauge structure of the low energy theory and another containing the tables relating the original lattice deformations with resulting gauge fields.

\section{Preliminaries}
We consider the tight-binding model in which the propagating  electron dynamics is  described by the two-dimensional hexagonal lattice Hubbard model, given by the Hamiltonian,
\begin{equation}\label{TBHamiltonian}
    H=-t\sum_{n,\hat{a}}\left(a^\dag_n b_{n+\hat{a}}+b^\dag_{n+\hat{a}}a_n\right),
\end{equation}
where $a^\dag_n$ and $a_n$ are the creation/annihilation operators for the electron on the A lattice site, and,  respectively $b^\dag_n$ and $b_n$ are those for the B lattice sites. Each A site is connected to three nearest neighbor B sites by lattice vectors $\hat{a}=\hat{1},\hat{2},\hat{3}$ (see the Fig.\ref{fig:LatticeVectors}),
\begin{equation}\label{LatticeVectors}
   \hat{1}=a(1,0),\quad
   \hat{2}=a(-1/2,\sqrt{3}/2),\quad
   \hat{3}=a(-1/2,-\sqrt{3}/2).
\end{equation}
and \emph{vice versa} each B site is connected to nearest A sites by lattice vectors $-\hat{a}$. 
\input{hexagonal.TpX}

The  lattice can be represented as a superposition of two simple (Bravais) lattices (A and B sublattices), each generated by lattice vectors $\{\hat{l}_1=\hat{1}-\hat{2}, \hat{l}_2=\hat{1}-\hat{3}\} $ (see the Fig. \ref{fig:LatticeVectors}),
\begin{equation}
    \hat{l}_1=a(\ft32,-\ft{\sqrt{3}}{2}),\quad
	\hat{l}_2=a(\ft32,\ft{\sqrt{3}}{2}).
\end{equation} 
The B-sublattice is  shifted by $\hat{1}$ with the respect to the A-sublattice. The spacing for the Bravais lattice is $|\hat{l}_1|=|\hat{l}_2|=\sqrt{3}a$.

The sublattice degeneracy has an important implication. Thus, a lattice  function $f(x_{\mathbf{n}})$, will be described by a two component object $f_{\alpha,\mathbf{n}}$, where $\alpha=A,B$ parameterizes the sublattice and $\mathbf{n}$, Bravais lattice site within the sublattice.

In the case of Hamiltonian \eqref{TBHamiltonian}, the electron's annihilation operators can also be combined into a two component field $\Psi_{\alpha,\mathbf{n}}=(a_{\mathbf{n}},b_{\mathbf{n}})$. In terms of the new field the Hamiltonian \eqref{TBHamiltonian} takes the following form,
\begin{multline}\label{TBHBz}
     H=-t\sum_{\mathbf{n}}\left(\Psi^\dag_{A,\mathbf{n}}\cdot \Psi_{B,\mathbf{n}}+
    \Psi^\dag_{A,\mathbf{n}}\cdot \Psi_{B,\mathbf{n}-\hat{l}_1}+
    \Psi^\dag_{A,\mathbf{n}}\cdot \Psi_{B,\mathbf{n}-\hat{l}_2}+\text{h.c.}\right)\\
    \equiv-t\sum_{\mathbf{n}} \Psi_{\mathbf{n}}^\dag\cdot D \cdot \Psi_{\mathbf{n}},
\end{multline}
where
\begin{equation}
    D=
   \begin{pmatrix}
   0 & 1+T^\dag_{\hat{l}_1}+T^\dag_{\hat{l}_2}\\
   1+T_{\hat{l}_1}+T_{\hat{l}_2} & 0
\end{pmatrix},
\end{equation}
where $T_{\hat{l}_i}$, $i=1,2$ represent elementary translation operators on the Bravais lattice.

Taking into account the anticommutation relations, in terms of the new fields,
\begin{equation}
    \{\Psi_{\alpha,\mathbf{n}},\Psi_{\beta,\mathbf{m}}^\dag\}=
    \delta_{\alpha \beta}\delta_{\mathbf{n},\mathbf{m}},
\end{equation}
we can write down the discrete Lagrangian corresponding to the Hamiltonian eq.\eqref{TBHFD},
\begin{multline}\label{handyAction}
   L=\ii\sum_{\alpha\mathbf{n}}
   \Psi^\dag_{\alpha,\mathbf{n}} \cdot \dot{\Psi}_{\alpha \mathbf{n}}-H(\Psi^\dag,\Psi)\\
   =\sum_{\mathbf{n}}\left[
   \ii\Psi^\dag_{A,\mathbf{n}} \cdot \dot\Psi_{A,\mathbf{n}}+
   \ii\Psi^\dag_{B,\mathbf{n}} \cdot \dot\Psi_{B,\mathbf{n}}\right.\\
  +\left.t
    \left(\Psi^\dag_{A,\mathbf{n}}\cdot \Psi_{B,\mathbf{n}}+
    \Psi^\dag_{A,\mathbf{n}}\cdot \Psi_{B,\mathbf{n}-\hat{l}_1}+
    \Psi^\dag_{A,\mathbf{n}}\cdot \Psi_{B,\mathbf{n}-\hat{l}_2}+\text{h.c.}\right)\right]
\end{multline}

\subsubsection*{The Low energy theory}
To find the low energy theory, let us go to the momentum space representation of the action \eqref{handyAction}, by using the Fourier transforms of the fields,
\begin{equation}\label{Fourier}
    \widetilde{\Psi}_{\alpha}(\mathbf{k})=\sum_{\mathbf{n}}\Psi_{\alpha,\mathbf{n}}\,\e^{\ii \mathbf{k}\cdot \mathbf{n}}
\end{equation}
where $\mathbf{k}=(k_x,k_y)$  is the lattice momentum.  This definition implies, that the Fourier transform $\widetilde{\Psi}_\alpha(\mathbf{k})$ is periodic with periods given by the reciprocal lattice vectors $\{\hat{k}_1,\hat{k}_2\}$ satisfying,
\begin{equation}\label{DualBasis}
    \hat{k}_i\cdot \hat{l}_j=2\pi\delta_{ij}.
\end{equation}
In our setup the reciprocal lattice vectors are given by
\begin{equation}
    \hat{k}_1=\ft{2\pi}{3a}(1,-\sqrt{3}),\quad \hat{k}_2=\ft{2\pi}{3a}(1,\sqrt{3}),
\end{equation}
and $|\hat{k}_1|=|\hat{k}_2|={4\pi}/{3a}$.

The inverse Fourier transform is given by,
\begin{equation}\label{FourierInv}
    \Psi_{\alpha,\mathbf{n}}=\ft{1}{A_{\mathrm{Bz}}} \int_{\rm FD}\dd^2k\, \Psi_{\alpha}(\mathbf{k})\e^{-\ii \mathbf{k}\cdot \mathbf{n}},
\end{equation}
where the integration is done over the fundamental domain (Brillouin zone) $-1/2\leq k_i <1/2$, $i=1,2$ and $A_{\mathrm{Bz}}=8\pi^2/3 \sqrt{3}a^2$ is the area of the Brillouin zone. 

 The action in terms of Fourier transforms takes the form,
\begin{equation}\label{actionFourier}
    S=\ft{1}{A_{\mathrm{Bz}}}
    \int_{\rm FD}\dd t\dd^2k\,
    \left[\ii\Psi^\dag(k) \dot{\Psi}(k)+t\Psi^\dag(k)\cdot D(k)\cdot \Psi(k)\right].
\end{equation}
where
\begin{equation}
   D(k)=
 \begin{pmatrix}
   0       & d(k) \\
   d^*(k)   & 0
 \end{pmatrix},
\end{equation}
is the fermionic operator, and
\begin{equation}
   d(k)=1+\e^{\ii\mathbf{k}\cdot \hat{l}_1}+\e^{\ii\mathbf{k}\cdot \hat{l}_2}
  .
\end{equation}

In the Brillouin zone the function $d(k)$ vanishes for two distinguished values of  $k$: $\mathbf{K}_1\equiv -\mathbf{K}=(0,-\ft{4 \pi \sqrt{3}}{9a}) $ and $\mathbf{K}_2\equiv \mathbf{K}=(0,\ft{4 \pi \sqrt{3}}{9a})$. These two points are called \emph{Dirac points}. Since the Hamiltonian vanishes at these points, their neighborhoods are relevant for the low energy regime of the model. In the hexagonal Brillouin zone picture the Dirac points appear at the corners of the zones. In the the rhombic Brillouin zone description, used by us, they appear in the core of the zone. 

As we restrict ourselves to the dynamics near the Dirac points, the leading contribution comes from the linear terms in the expansion of the fermionic operator, which is controlled by the expansion of the function $d(\mathbf{k})$,
\begin{equation}
    d(\pm \mathbf{K}+\mathbf{k})
   = \ft{3a}{2}(-\ii k_x\pm \ii (-\ii k_y))+\dots \equiv
   \mathbf{k}\cdot \nabla(\pm \mathbf{K})+\dots,
\end{equation}
where $\nabla(\pm \mathbf{k})$ is the momentum space gradient of $d(\mathbf{k})$.
Using this expansion we can express the fermionic matrix $D(k)$ in the following form,
\begin{equation}
    D(\pm K+k)=\ft{3a}{2}\left[(-\ii k_x) \sigma_2\mp (-\ii k_y)\sigma_1\right]+\dots
\end{equation}
The $\pm$ sign labeling the Dirac point. If we introduce an additional index labeling the Dirac point, then we can rewrite the fermionic operator in the leading order in the following form,
\begin{equation}
    D(k)=\ft{3a\ii}{2}\left[(-\ii k_x) \beta+ (-\ii k_y) \rho	\right]
\end{equation}
where we introduced the matrices
\begin{equation}\label{beta-rho}
    \beta=\sigma_2\otimes \I,\quad \rho=-\sigma_1\otimes \sigma_3.
\end{equation}

If we chose $\gamma^0$ matrix as,
\begin{equation}\label{gamma0}
    \gamma^0=\ii \sigma_3\otimes \sigma_3.
\end{equation}
and introduce the other two Dirac gamma matrices as, 
\begin{equation}\label{gamma12}
    \gamma^1=- \gamma^0\beta=
    -\sigma_1   \otimes \sigma_3,
\quad 
     \gamma^2=- \gamma^0\rho
     =-\sigma_2\otimes \I,
\end{equation}
as well as the Dirac conjugate of the wave function,
\begin{equation}
   \bar{\Psi}=\Psi^\dag\cdot \gamma^0,
\end{equation}
the action \eqref{actionFourier} in low energy rewrites as,
\begin{equation}\label{actionFourierFree}
    S=\ft{1}{A_{\mathrm{FP}}}\int_{ \sim\rm Dp}\dd k_x\dd k_y\, 
   \left[ \ii \bar\Psi(k) \gamma_0\dot{\Psi}(k)
    +\ft{3ta}{2}
    \ii \bar\Psi(k) (-\ii k_m)\gamma_m\Psi(k)\right],
\end{equation}
where the index $m=x,y$, and integration is performed in the vicinity of the Dirac points assuming some implicit energy cut-off.

Representing now the low energy field as a Fourier transform of a continuous function $\Psi(x)$,
\begin{equation}
   \Psi(\mathbf{k})=\frac{\sqrt{A_{\mathrm{FP}}}}{2\pi}\int \dd^2 \mathbf{r}\, \Psi(\mathbf{r})\e^{\ii \mathbf{k}\cdot \mathbf{r}},
\end{equation}
where $\mathbf{k}$ is treated as an ordinary continuum space momentum, the action becomes,
\begin{equation}\label{LEFree}
   S=\int\dd t\dd^2 x\,
   (\ii \bar{\Psi}\gamma^0 \pd_0 \Psi+\ii v_{\mathrm{F}}\bar{\Psi}\gamma^i \pd_i \Psi),
\end{equation}
where $i=1,2$  and $v_{\mathrm{F}}=3ta/2$ is the Fermi velocity, which plays here the role of the speed of light. The Dirac Fermi field $\Psi$ carries one index $\alpha=A,B$ labeling sublattice, one index $\pm$ denoting the Dirac point, in total four components. Here one can include also the electron spin index, which should rise the number of components up to eight, but we are not considering the spin component in this work.

 
\section{Defects and Hubbard model deformations}
So far we considered the two-dimensional lattice as a regular rigid object.  In the real world, however, the lattice is a  subject to smooth deformations as well as to discontinuous topological defects and impurities. 

The smooth deformations correspond to phonon fields. In the standard approach, the basic variable describing the phonon field is the position of the individual atoms in the lattice (see e.g. \cite{SaitoDressekhaus2x:PhProp}). Then, in the linear response theory the phonon dynamics can be described by using the dynamical matrix encoding the elastic forces between the atoms. In the case of graphene is well established that the dynamical matrix should involve elastic forces for at least up to next-to-neighbor atom site, due to relative small difference in the distances to nearest and next-to-nearest neighbors.

In the framework of the microscopic tight-binding model, the coupling to the fermionic field is realized through the point dependent modification of the hopping amplitudes in the Hubbard model. For small deformations the amplitude modifications are linear in the displacements of atoms from the equilibrium position. In this approximation, the description of the phonon field in terms of the displacements is equivalent to the description in terms of transition amplitude variations.  For large deformations this may not remain true, however, the  description in terms of amplitudes is advantaged because the amplitudes are always linearly coupled to the electronic field. In the low energy theory such deformations were shown to give rise to an interaction potential similar to gauge coupling \cite{PhysRevLett.78.1932}. 

On the other hand the study of various topological defects was shown to lead to the interaction potentials having the form of coupling to nonabelian gauge field backgrounds \cite{PhysRevLett.69.172,Gonzalez1993771,PhysRevLett.85.5190,springerlink:10.1134-1.1387528,Furtado20085368}.
The distinguishing feature of a topological defect is that, in contrast to the phonon field, the `defects field' may not necessarily have a limit as a continuous function in the low energy theory. This property of topological defects plays a crucial role in building up the \emph{nonabelian} gauge field, since for coupling electronic modes at different Dirac points can be only achieved through discontinuous modes. 

All in all, both phonon fields and lattice defects can be taken into account in the tight-binding approach through the deformation of the Hubbard model Hamiltonian by adding to it certain types of operators. In what follows we will consider such deformations involving the local, nearest neighbor and next-to-nearest neighbor terms. Strictly speaking, certain types of topological defects should involve non-local terms of arbitrary long range but, as we can anticipate, the terms only up to the next-to leading are enough to generate a generic gauge field.

\subsection{Hubbard model deformations}

Let us turn to the tight-binding model described by eq.\eqref{TBHamiltonian}, and write down possible deformation terms. 

Consider first the local contribution. A generic lattice deformation may lead to local modification of the Fermi level. This modification can be taken into account by the following terms added to the Hamiltonian,
\begin{equation}
   \Delta H_{\mathrm{n}}=-\sum_{\mathbf{n}}(a_{\mathbf{n}}^\dag z_{A\mathbf{n}}a_{\mathbf{n}}+
   b_{\mathbf{n}}^\dag z_{B\mathbf{n}}b_{\mathbf{n}}),
\end{equation}
with $z_{\alpha\mathbf{n}}$, $\alpha=A,B$ are the components of a real lattice function.

The nearest neighbor transition amplitude can also be modified. In general the amplitude can be complex, however, the corresponding term in the Hamiltonian should be Hermitian. Therefore, in the most general form it reads,
\begin{equation}
   \Delta H_{\mathrm{nn}}=-\sum_{\mathbf{n},\hat{a}}\left(
    a_{\mathbf{n}}^\dag z_{\mathbf{n},\hat{a}} b_{\mathbf{n}+\hat{a}}+
    b_{\mathbf{n}+\hat{a}}^\dag \bar{z}_{\mathbf{n},\hat{a}} a_{\mathbf{n}}
    \right),
\end{equation}
the bar stands for the complex conjugate.

 In addition to the nearest hopping the deformation, one can induce the next-to-nearest one. While for the undeformed model the amplitude of such hopping is expected to be much smaller than the nearest neighbor counterpart, its variation due to the lattice deformation could be rather large. This deformation is taken into account by adding the following term to the Hamiltonian,
\begin{equation}
	\Delta H_{\mathrm{nnn}}=-\sum_{\mathbf{n},\hat{b}\neq\hat{a}}\left(
	a_{\mathbf{n}}^\dag z_{A\mathbf{n},\hat{a}\hat{b}} a_{\mathbf{n}+\hat{a}-\hat{b}}
	+b_{\mathbf{n}-\hat{a}}^\dag z_{B\mathbf{n},\hat{a}\hat{b}} b_{\mathbf{n}-\hat{b}}\right),
\end{equation}
where $ z_{\mathbf{n},\hat{a}\hat{b}}=\bar{z}_{\mathbf{n},\hat{b}\hat{a}}$, $\alpha=A,B$ are the transition amplitudes for the hopping between nearest A-sites and nearest B-sites respectively.

Thus, in total we have three types of parameters describing the lattice deformation: real valued site based (scalar) field $z_{\mathbf{n}}$, link based (vector) field $z_{\mathbf{n},\hat{a}}$ and face based (2D-pseudoscalar) field $z_{\mathbf{n},\hat{a}\hat{b}}$.

Tout ensemble, the above modifications lead to the following deformed Lagrangian \eqref{handyAction},
\begin{multline}\label{FullLagrangian}
   L=L_0-\Delta H_{\mathrm{n}}-\Delta H_{\mathrm{nn}}-\Delta H_{\mathrm{nnn}}\\
	=L_0+
	\sum_{\mathbf{n}}\left[
	\Psi_{A,\mathbf{n}}^\dag z_{A,\mathbf{n}} \Psi_{A,\mathbf{n}}
     +\Psi_{B,\mathbf{n}}^\dag z_{B,\mathbf{n}} \Psi_{B,\mathbf{n}}
     \right.~~~~~~~~~~~~~~~\\
  +
    \left(\Psi^\dag_{A,\mathbf{n}} z_{\mathbf{n},\hat{1}} \Psi_{B,\mathbf{n}}+
    \Psi^\dag_{A,\mathbf{n}}z_{\mathbf{n},\hat{2}} \Psi_{B,\mathbf{n}-\hat{l}_1}+
    \Psi^\dag_{A,\mathbf{n}}z_{\mathbf{n},\hat{3}} \Psi_{B,\mathbf{n}-\hat{l}_2}+\text{h.c.}\right)\\
	+
	\left(
	\Psi^\dag_{A,\mathbf{n}}  z_{A \mathbf{n},\hat{1}\hat{2}} \Psi_{A,\mathbf{n}-\hat{l}_1}
	+\Psi^\dag_{A,\mathbf{n}} z_{A \mathbf{n},\hat{1}\hat{3}} \Psi_{A,\mathbf{n}-\hat{l}_2}
	+\Psi^\dag_{A,\mathbf{n}}  z_{A \mathbf{n},\hat{2}\hat{3}} \Psi_{A,\mathbf{n}+\hat{l}_1-\hat{l}_2}
	+\text{h.c.}	\right)\\
	+\left.\left(
	\Psi^\dag_{B,\mathbf{n}}  z_{B \mathbf{n},\hat{1}\hat{2}} \Psi_{B,\mathbf{n}-\hat{l}_1}
	+\Psi^\dag_{B,\mathbf{n}} z_{B \mathbf{n},\hat{1}\hat{3}} \Psi_{B,\mathbf{n}-\hat{l}_2}
	+\Psi^\dag_{B,\mathbf{n}}  z_{B \mathbf{n},\hat{2}\hat{3}} \Psi_{B,\mathbf{n}+\hat{l}_1-\hat{l}_2}
	+\text{h.c.}	\right)
	\right],
\end{multline}
or in the compact form,
\begin{multline}\label{compacLagrangian}
	L=L_0+\sum_{\mathbf{n}}\Psi_\mathbf{n}^\dag\cdot Z_{\mathbf{n}} \cdot \Psi_{\mathbf{n}} 
	+\sum_{\mathbf{n},i=1,2}(\Psi_\mathbf{n}^\dag\cdot
	 Z_{\mathbf{n},\hat{l}_i} \cdot 
	\Psi_{\mathbf{n}-\hat{l}_i}+
	\Psi_{\mathbf{n}-\hat{l}_i}^\dag\cdot
	Z^{*}_{\mathbf{n},\hat{l}_i} \cdot 
	\Psi_{\mathbf{n}} ) \\
	+\sum_{\mathbf{n}}(
	\Psi^\dag_{\mathbf{n}-\hat{l}_1}\cdot 				
	Z_{\mathbf{n},\hat{l}_1\hat{l}_2} \cdot \Psi_{\mathbf{n}-\hat{l}_2}
	+\Psi^\dag_{\mathbf{n}-\hat{l}_2}\cdot 				
	Z^{*}_{\mathbf{n},\hat{l}_1\hat{l}_2} \cdot \Psi_{\mathbf{n}-\hat{l}_1}),
\end{multline}
where we use the capitalized symbols to denote the following $2 \times 2$ matrices,
\begin{align}\label{z-fields}
	&Z_{\mathbf{n}}	=
	\ft{1}{2}(z_{A,\mathbf{n}}+z_{B,\mathbf{n}})\I
	+\ft{1}{2}(z_{A,\mathbf{n}}-z_{B,\mathbf{n}}) \sigma_3
	+z_{\mathbf{n},\hat{1} }\ft{1}{2}( \sigma_1+\ii \sigma_2)+\bar{z}_{\mathbf{n},\hat{1} }\ft{1}{2} (\sigma_1-\ii \sigma_2)\nonumber \\
	&\qquad \equiv z_{\mathbf{n}}^0 \I+z_{\mathbf{n}}^i \sigma_i,\quad
	\bar{z}^0_{\mathbf{n}}=z^0_{\mathbf{n}},\quad \bar{z}^i_{\mathbf{n}}=z^i_{\mathbf{n}}\\	
	&Z_{\mathbf{n},\hat{l}_i}=z_{\mathbf{n},\hat{l}_i}\sigma_{+}
 	+z^0_{\mathbf{n},\hat{l}_i}\I+z^3_{\mathbf{n},\hat{l}_i} \sigma_3,\quad
	Z^{*}_{\mathbf{n},\hat{l}_i}=\bar{z}_{\mathbf{n},\hat{l}_i}\sigma_{-}
 	+\bar{z}^0_{\mathbf{n},\hat{l}_i}\I+\bar{z}^3_{\mathbf{n},\hat{l}_i} \sigma_3,\\
	&Z_{\mathbf{n},\hat{l}_1 \hat{l}_2}=
	z^0_{\mathbf{n},\hat{l}_1 \hat{l}_2}\I
	+	z^3_{\mathbf{n},\hat{l}_1 \hat{l}_2} \sigma_3,\quad
	Z^{*}_{\mathbf{n},\hat{l}_1 \hat{l}_2}=
	\bar{z}^0_{\mathbf{n},\hat{l}_1 \hat{l}_2}\I
	+	\bar{z}^3_{\mathbf{n},\hat{l}_1 \hat{l}_2} \sigma_3,
\intertext{with}
	&z_{\mathbf{n},\hat{l}_1}=z_{\mathbf{n},\hat{2} },\quad 
	z_{\mathbf{n},\hat{l}_2}=z_{\mathbf{n},\hat{3} },\quad
	\bar{z}_{\mathbf{n},\hat{l}_1}=\bar{z}_{\mathbf{n},\hat{2} },\quad 
	\bar{z}_{\mathbf{n},\hat{l}_2}=\bar{z}_{\mathbf{n},\hat{3} },\\
	&z^0_{\mathbf{n},\hat{l}_1}=
     \ft{1}{2}(z_{A\mathbf{n},\hat{1}\hat{2}}+z_{B\mathbf{n},\hat{1}\hat{2}}) ,\quad 
	z^0_{\mathbf{n},\hat{l}_2}=
	\ft{1}{2}(z_{A\mathbf{n},\hat{1}\hat{3}}+z_{B\mathbf{n},\hat{1}\hat{3}}),\\
	&z^3_{\mathbf{n},\hat{l}_1}=
	 \ft{1}{2}(z_{A\mathbf{n},\hat{1}\hat{2}}-z_{B\mathbf{n},\hat{1}\hat{2}}),\quad
	z^3_{\mathbf{n},\hat{l}_2}=
	\ft{1}{2}(z_{A\mathbf{n},\hat{1}\hat{3}}-z_{B\mathbf{n},\hat{1}\hat{3}}),\\
	&z^0_{\mathbf{n}\hat{l}_1 \hat{l}_2}=
	\ft{1}{2}(z_{A\mathbf{n},\hat{2}\hat{3}}+z_{B\mathbf{n}\hat{2}\hat{3}}),\quad 				
	z^3_{\mathbf{n},\hat{l}_1 \hat{l}_2}=
	\ft{1}{2}(z_{A\mathbf{n},\hat{2}\hat{3}}-z_{B\mathbf{n},\hat{2}\hat{3}}).
\end{align}

The Pauli matrices $\sigma_{1,2}$ and $\sigma_3$, as well as unity matrix $\I$ are defined in the sublattice space.

The `lattice vector' $Z_{\mathbf{n},\hat{l}_i}$ appearing in the Lagrangian \eqref{compacLagrangian}, can be expressed as a scalar product of a local vector field $\hat{\mathbf{Z}}_{\mathbf{n}}$ and lattice vector $\hat{l}_i$:
\begin{equation}
	Z_{\mathbf{n},\hat{l}_i}=\mathbf{Z}_{\mathbf{n}}\cdot \hat{l}_i.
\end{equation}
The inverse transformation reads,
\begin{equation}
	\mathbf{Z}_{\mathbf{n}}=\ft{1}{2\pi} \sum_{i} Z_{\mathbf{n},\hat{l}_i} \hat{k}_i, 
\end{equation}
where $\hat{k}_i$, $i=1,2$ are the vectors of the dual basis given by eq.\eqref{DualBasis}.

\subsection{The low energy limit}

Let us now take the low energy limit of the theory described by the Lagrangian \eqref{compacLagrangian} and find the coupling of the low energy electronic modes the modes of fields introduced in the previous section. In order to do this, let us again expand the action near the Dirac points for the fermionic field and keep only those modes of the deformation fields which couple to the low energy modes of the fermion. In the momentum space the expanded action reads,
\begin{multline}\label{LELang}
    S=S_0+\ft{1}{A_{\mathrm{Bz}}^2} \int \dd^2k\dd^2q
    \bigl[
    \Psi_{+}^\dag(k)Z (k-q) \Psi_{+}(q)
   +\Psi_{-}^\dag(k) Z(k-q) \Psi_{-}(q)\\
   + \Psi_{+}^\dag(k)Z_{-}(k-q) \Psi_{-}(q)
   + \Psi_{-}^\dag(k)Z_{+}(k-q) \Psi_{+}(q)\bigr]\\
   -\ii \bigl[
   \Psi_{+}^\dag (k)\bigl\{{\mathbf{Z}}(k-q)\cdot \nabla_{+}
   +\mathbf{Z}^{*}(k-q)\cdot \nabla_{-}\bigr\}\Psi_{+}(q) \\
   +  \Psi_{-}^\dag (k)\bigl\{{\mathbf{Z}}(k-q)\cdot \nabla_{-}
   +{\mathbf{Z}}^{*}(k-q)\cdot \nabla_{+}\bigr\}\Psi_{-}(q) \\
   +\Psi_{+}^\dag (k)\bigl\{{\mathbf{Z}}_{-}(k-q)\cdot \nabla_{-}
   +{\mathbf{Z}}^{*}_{-}(k-q)\cdot \nabla_{-}\bigr\}\Psi_{-}(q) \\
   + \Psi_{-}^\dag (k)\bigl\{ {\mathbf{Z}}_{+}(k-q)\cdot \nabla_{+}
   +{\mathbf{Z}}^{*}_{+}(k-q)\cdot \nabla_{+}\bigr\}\Psi_{+}(q)\bigr]\\
	+ \Psi_{+}^\dag(k)\bigl\{\e^{2\pi\ii/3}Z_{\hat{l}_1\hat{l}_2}(k-q)
	+\e^{-2\pi\ii/3}Z^{*}_{\hat{l}_1\hat{l}_2}(k-q)\bigr\} \Psi_{+}(q)\\
   +\Psi_{-}^\dag(k) \bigl\{\e^{-2\pi\ii/3}Z_{\hat{l}_1\hat{l}_2}(k-q)
	+\e^{2\pi\ii/3}Z^{*}_{\hat{l}_1\hat{l}_2}(k-q)\bigr\} \Psi_{-}(q)\\
   + \Psi_{+}^\dag(k)\bigl\{Z_{\hat{l}_1\hat{l}_2-}(k-q)
	+Z^{*}_{\hat{l}_1\hat{l}_2-}(k-q)\bigr\} \Psi_{-}(q)\\
   + \Psi_{-}^\dag(k)\bigl\{Z_{\hat{l}_1\hat{l}_2+}(k-q)
	+Z^{*}_{\hat{l}_1\hat{l}_2+}(k-q)\bigr\} \Psi_{+}(q)
,
\end{multline}
where the subscripts $\pm$ of $Z_{\pm}$,  $\mathbf{Z}_{\pm}$ and $Z_{\hat{l}_1,\hat{l}_2 \pm }$  indicate, that the field is evaluated near the respective Dirac point, e.g.,
\begin{equation}
     \mathbf{Z}_{\pm}(\mathbf{k}) \equiv\mathbf{Z}(\pm \mathbf{K}+\mathbf{k}),
\end{equation}
absence of such a subscript means that the field is taken near the origin of momentum space. We also included the multiplicative coefficient into the momentum measure.

In the matrix form the Lagrangian \eqref{LELang} takes the form,\footnote{We are using the first factor for the sublattice space, while the second factor corresponds to the Dirac point space.}
\begin{equation}
	L=L_0+ Z^{IJ}\Psi^\dag\cdot\sigma_I \otimes \sigma_J \cdot \Psi,
\end{equation}
where $I,J=0,1,2,3$ and $\sigma_0=\I$, while the other sigma matrices are Pauli matrices. The field $Z^{IJ}$ consists of three main contributions:
\begin{equation}\label{final-non-standard}
	Z^{IJ}=Z^{IJ}_{\mathrm{n}}+Z^{IJ}_{\mathrm{nn}}+Z^{IJ}_{\mathrm{nnn}},
\end{equation}
where $Z^{IJ}_{\mathrm{n}}$ comes from the local deformation of the Hamiltonian, $Z^{IJ}_{\mathrm{nn}}$ from the nearest neighbor and $Z^{IJ}_{\mathrm{nnn}}$ from the next to nearest neighbor deformations described respectively by terms $\Delta H_{\mathrm{n}}$, $\Delta H_{\mathrm{nn}}$ and $\Delta H_{\mathrm{n}}$ in the Hamiltonian. The tables of the values of  $Z^{IJ}_{\mathrm{n,nn,nnn}}$ in terms of the original deformation parameters are given in the appendix~\ref{appendix:tables}.


Replacing the Hermitian conjugate wave function by the Dirac conjugate according to  $\bar{\Psi}=\Psi^\dag \gamma^0$,  the Lagrangian \eqref{final-non-standard} becomes,
\begin{equation}\label{final-coupling}
	L=L_0+\Phi \bar{\Psi}\Psi+U^a\bar{\Psi}\tau_a  \Psi+ A_\mu \bar{\Psi}\gamma^\mu  \Psi+
	 B^a_{\mu}\bar{\Psi}\gamma^\mu \tau_a \Psi,
\end{equation}
where the field $\Phi$ and $U=U^a \tau_a$ are Abelian scalar and su(2) algebra-valued scalar fields, while $A_\mu$ and $B_\mu=B^a_\mu \tau_a$ are, respectively, Abelian gauge (pseudo-electromegnetic) field and $B_\mu$ su(2) non-abelian gauge fields. The effective fields are expressed in terms of the original deformation modes as follows,
\begin{subequations}\label{final-fields}
\begin{align}
	\Phi&=-\ft{3}{4}(z_{y}^3-\bar{z}_{y}^3)
	+\ft{\sqrt{3}}{4}(z^{\prime 3}-\bar{z}^{\prime 3})  ,
	\\ \nonumber
	U^1&=-\ft{1}{4}(z_{\hat{1}-}+\bar{z}_{\hat{1}-}-z_{\hat{1}+}-\bar{z}_{\hat{1}+})\\
	&-\ft{3}{8}[z_{x-}+\bar{z}_{x-}-z_{x+}-\bar{z}_{x+}
	-\ii (z_{y+}+\bar{z}_{y+}+z_{y-}+\bar{z}_{y-})],
	\\ \nonumber
	U^2&=\ft{\ii}{4}(z_{\hat{1}-}+\bar{z}_{\hat{1}-}+z_{\hat{1}+}+\bar{z}_{\hat{1}+})\\
	&-\ft{3\ii}{8}[z_{x-}+\bar{z}_{x-}+z_{x+}+\bar{z}_{x+}
	+\ii (z_{y+}+\bar{z}_{y+} -z_{y-}-\bar{z}_{y-})],\\
	U^3&=-\ft{\ii}{2}(z_A-z_B)+\ft{3\ii}{4}(z^3_{x}+\bar{z}^3_{x})
	+\ft{\ii}{4}(z^{\prime 3}+\bar{z}^{\prime 3})  ,
\end{align}
for the scalar fields,
\begin{align}
	A_0&=-\ft{1}{2}(z_A+z_B)+\ft{3}{4}(z^0_x+\bar{z}^0_x)-\ft{1}{4}(z^{\prime 0}_x+\bar{z}^{\prime 0}_x),\\ 
	A_1&=\ft{\ii}{2}(z_{\hat{1}}-\bar{z}_{\hat{1}})-\ft{3\ii}{8} (z_x-\bar{z}_x) ,\\ 
	A_2&=-\ft{3\ii}{8}(z_{y}-\bar{z}_{y}),
\end{align}
for the pseudo-electromagnetic field and, finally,
\begin{align}
\nonumber
      B^1_0&=-\ft{\ii}{4}(z_{\hat{1}-}-\bar{z}_{\hat{1}-}+z_{\hat{1}+}-\bar{z}_{\hat{1}+})\\
	&+\ft{3\ii}{8}[z_{x-}-\bar{z}_{x-}+z_{x+}-\bar{z}_{x+}
	+\ii (z_{y+}-\bar{z}_{y+}-z_{y-}+\bar{z}_{y-})],\\ \nonumber
	B^1_1&=\ft{1}{4}(z_{A+}+z_{A-}+z_{B+}+z_{B-})\\ \nonumber
	-\ft{3}{4}&[z_{x-}^0+\bar{z}_{x-}^0+z_{x+}^0+\bar{z}_{x+}^0
	+\ii(z_{y+}^0+\bar{z}_{y+}^0-z_{y-}^0-\bar{z}_{y-}^0)]\\
	&~~~~+\ft{1}{4}(z_{-}^{\prime 0} +z_{+}^{\prime 0}+\bar{z}_{-}^{\prime 0}+\bar{z}_{+}^{\prime 0})\\ \nonumber
	B^1_2&= \ft{\ii}{4}(z_{A-}-z_{B-}-z_{A+}+z_{B+}) \\ \nonumber
     & -\ft{3\ii}{4}[z_{x-}^3+\bar{z}_{x-}^3-z_{x+}^3-\bar{z}_{x+}^3
	-\ii(z_{y+}^3+\bar{z}_{y+}^3+z_{y-}^3+\bar{z}_{y-}^3)]\\ 
	&+\ft{\ii}{4}(z_{-}^{\prime 3} +z_{+}^{\prime 3}-\bar{z}_{-}^{\prime 3}-\bar{z}_{+}^{\prime 3}),
\end{align}
\begin{align}
 \nonumber
	B^2_0&=\ft{1}{4}(z_{\hat{1}-}-\bar{z}_{\hat{1}-}-z_{\hat{1}+ }+\bar{z}_{\hat{1}+ } )\\
	&-\ft{3}{8}[z_{x-}-\bar{z}_{x-}-z_{x+}+\bar{z}_{x+}
	-\ii (z_{y+}-\bar{z}_{y+}+z_{y-}-\bar{z}_{y-})],\\ \nonumber
	B^2_1&= \ft{\ii}{4}(z_{A-} -z_{A+}+z_{B-}-z_{B+})\\  \nonumber
	&-\ft{3\ii}{4}[z_{x-}^0+\bar{z}_{x-}^0-z_{x+}^0-\bar{z}_{x+}^0
	-\ii(z_{y+}^0+\bar{z}_{y+}^0+z_{y-}^0+\bar{z}_{y-}^0)]\\
	&	+\ft{\ii}{4}(z_{-}^{\prime 0} +z_{+}^{\prime 0}-\bar{z}_{-}^{\prime 0}-\bar{z}_{+}^{\prime 0}),\\ 
\nonumber
	B^3_2&=-\ft{1}{4}(z_{A-}-z_{B-}+z_{A+}-z_{B+})\\ \nonumber
	&+\ft{3}{4}[z_{x-}^3+\bar{z}_{x-}^3+z_{x+}^3+\bar{z}_{x+}^3
	+\ii(z_{y+}^3+\bar{z}_{y+}^3-z_{y-}^3-\bar{z}_{y-}^3)]\\
	&+\ft{1}{4}(z_{-}^{\prime 3} +z_{+}^{\prime 3}+\bar{z}_{-}^{\prime 3}+\bar{z}_{+}^{\prime 3}).
\end{align}
\begin{align}
	B^3_0&=\ft{3\ii}{4}(z_{y}^0-\bar{z}_{y}^0)
	-\ft{\ii \sqrt{3}}{4}(z^{\prime 0}-\bar{z}^{\prime 0}) ,
	\\
	B^3_1&=\ft{3}{8}(z_{y}+\bar{z}_{y}),\\ 
	B^3_2&=-\ft{1}{2}(z_{\hat{1}}+\bar{z}_{\hat{1}})+\ft{3}{8} (z_x+\bar{z}_x) ,
\end{align}
\end{subequations}
for the nonabelian gauge field. The shorthand notations are explained in the Appendix \ref{appendix:tables}.

\section{Example: The Stone-Wales defect}\label{sec:SWD}
\begin{figure}
\centering
\ifpdf
  \setlength{\unitlength}{1bp}%
  \begin{picture}(365.44, 116.21)(0,0)
  \put(0,0){\includegraphics{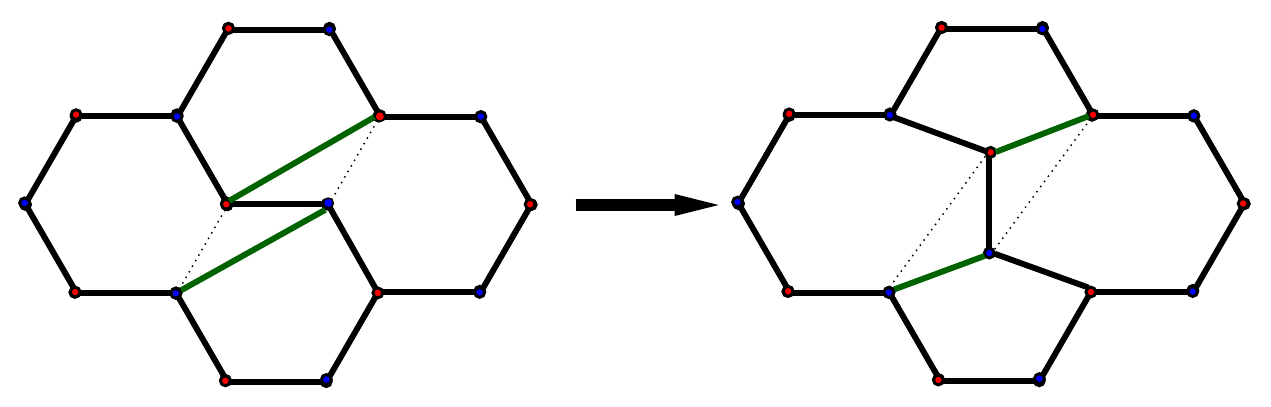}}
  \end{picture}%
\else
  \setlength{\unitlength}{1bp}%
  \begin{picture}(365.44, 116.21)(0,0)
  \put(0,0){\includegraphics{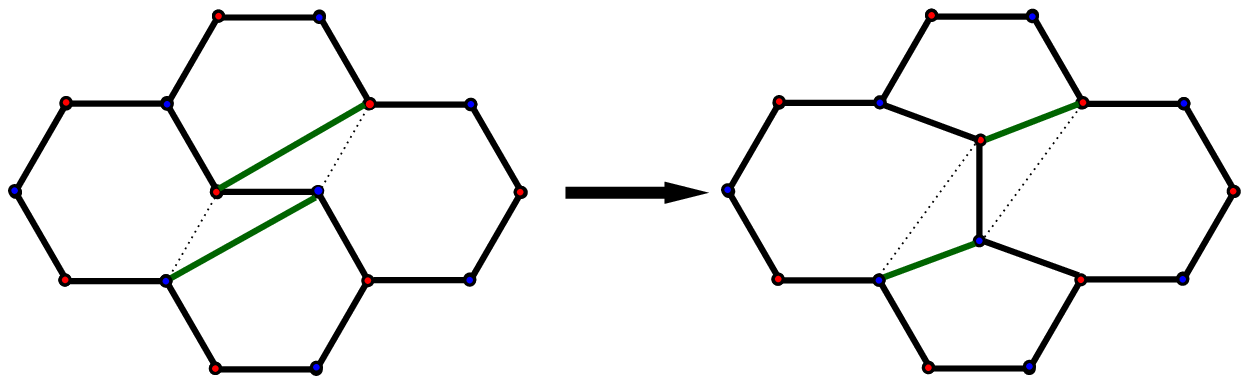}}
  \end{picture}%
\fi
\caption{\label{fig:Stone-Wales}%
 The Stone--Wales topological defect as a flip of two bonds adjacent to our unit cell bond. The flipped bonds are of the same orientation. Choosing another pair  of flipped bonds  results in an equivalent configuration with A and B atoms in the turning bond exchanging the roles. There are three different possible orientations for  Stone--Wales defect, corresponding to affected bond $\hat{1}$,  $\hat{2}$ or  $\hat{3}$.}
\end{figure}

As an example let us consider the Stone-Wales defect \cite{Stone1986501}. This defect consists of a couple of collateral heptagons paired by a couple of pentagons. As pentagons and heptagons are carrying opposite curvature `charges', this defect appears like a two-dimensional gravitational vortex. As was found by microscopic simulations, this type of vortices are  likely to be formed under strain as it relieves the stress in the direction of heptagons \cite{PhysRevB.57.R4277}. The defect can be obtained from a regular hexagonal lattice by flipping two links adjacent to a chosen (central) bond, as shown in the Fig.\ref{fig:Stone-Wales}. As there are three inequivalent types of links connecting the neighbor A and B sites, there are three different orientations of the defect. Let us denote each orientation as type $\hat{1}$, $\hat{2}$ and $\hat{3}$ according to the orientation of the central bond of the defect.

The Fig.\ref{fig:Stone-Wales} shows only the defect obtained from the alteration of a bond parallel to the unit cell, the other types are obtained through rotations by $\pm 120^o$ of the shown one. 
Although, the differently oriented defects differ by just rotation of the hexagonal lattice, in the compact notations they have quite a different form. Apart from the small smooth part, the defect field can be represented in the Hamiltonian by  nearest neighbor (nn) values $+t$ corresponding to the two canceled dotted bonds and two next to nearest (nnn) values $-t'\approx -t$ corresponding to the new emergent bonds.

For definitiveness let us consider only the type $\hat{1}$ defect, drawn in the Fig.\ref{fig:Stone-Wales}. From the picture it is not difficult to deduce that the defect field is represented by the following (sub)lattice deformation field,
\begin{equation}
	Z_{\mathbf{n},\hat{l}_1}=0,\quad 
	Z_{\mathbf{n},\hat{l}_2}=
\delta_{\mathbf{n},\mathbf{n}_0} t
\begin{pmatrix}
	0 & -1 \\
	0 & 1
\end{pmatrix}
+\delta_{\mathbf{n},\mathbf{n}_0+\hat{l}_2} t
\begin{pmatrix}
	1 &-1 \\
	0 & 0
\end{pmatrix}
,
\end{equation}
where $\mathbf{n}_0$ is the location of the turned bond, which place the role of the center of the defect. Using the low energy expansion worked out in the previous section we find that, as a result of this defect the electronic wave function is coupled to the following nonzero fields: The Yukawa scalar field,
\begin{subequations}\label{SWDfinal}
\begin{equation}
	\Phi=\ii g\left[-\cos(\mathbf{K}\cdot \mathbf{n}_0+\pi/6)\tau_1
	-\sin (\mathbf{K}\cdot \mathbf{n}_0+\pi/6) \tau_2+\ft{3}{2}\tau_3 
	\right]\delta^{(2)}(\mathbf{r}-\mathbf{n}_0),
\end{equation}
the Abelian gauge field given by,
\begin{equation}
	A_\mu=g (1/2,0,0)\delta^{(2)}(\mathbf{r}-\mathbf{n}_0),
\end{equation}
as well as nonabelian gauge field, 
\begin{multline}
	B_\mu = g \bigl(0,
	\sin(\mathbf{K}\cdot \mathbf{n}_0+\pi/6)\tau_1
	-\cos(\mathbf{K}\cdot \mathbf{n}_0+\pi/6)\tau_2
	+(\sqrt{3}/2) \tau_3,\\
	-\sqrt{3}\sin(\mathbf{K}\cdot \mathbf{n}_0+\pi/6)\tau_1
	+\sqrt{3}\cos(\mathbf{K}\cdot \mathbf{n}_0+\pi/6)\tau_2
	+(1/2) \tau_3\bigr)\\
	\times\delta^{(2)}(\mathbf{r}-\mathbf{n}_0),
\end{multline}
\end{subequations}
where $g=3\sqrt{3}ta^2/8\pi^2=\ft{\sqrt{3}}{4\pi^2} v_{\mathrm{F}}a$.

As can be deduced from the value of the parameter $g$, a single defect generates vanishingly weak effective field. Therefore, a sizable effect is reached, when a finite density of Stone--Wales defects is considered. 

On the other hand, as the explicit form Eq.\eqref{SWDfinal} shows up, the gauge and scalar depend discontinuously on the position of the defect, even for a fixed orientation. Indeed, since $\mathbf{K}\cdot \mathbf{n}_0=\ft{2\pi}{3}(n_{01}-n_{02})$, where $n_{0i}$, for $i=1,2$ are the integer lattice numbers of the defect position,  there are three distinct values for the gauge and Yukawa field for this orientation of the defect, depending on $(n_{01}-n_{02})=0,1,2 \mod 3$. A random distribution of Stone--Wales defects will result in the fields $\Phi$ and $B_\mu$ with vanishing average $\tau_1$ and $\tau_2$ components, but non-trivial Abelian gauge field and $\tau_3$ components. Different results one obtains when the defects form a sublattice with fixed $\theta=\mathbf{K}\cdot \mathbf{n}$.

\section{Conclusion}
In this work we considered deformations of the hexagonal lattice Hubbard model describing the graphene's electronic wave function in the tight-binding approximation. These deformations involved arbitrary space and time dependent operators including local, nearest neighbor and next-to-nearest neighbor interactions. The low energy effect of the deformation is the Yukawa and gauge couplings to the electronic Dirac field. The low energy Lagrangian is given by the eq. \eqref{final-coupling} and the fields are listed in eq. \eqref{final-fields}. There is no gravity-type coupling in this model since there is no spin connection in $2+1$ dimensional Clifford algebra.

The considered deformation operators do not exhaust all possible deformations in general, since one may include arbitrarily long-range interactions, however, they are enough to saturate generic gauge coupling: It is easy to check, e.g. by an inspection of the Tables \ref{table:local}, \ref{table:nearest} and \ref{table:next}, that excluding one of these contributions would result in an incomplete gauge field. On the other hand including longer range interaction will not add any new terms to the low energy action, but merely redefine (renormalize) the existent ones. Although the way the gauge fields are redefined is an interesting topic to study, in particular because there are topological defects which involve long range deformations.

The emergence of the whole gauge coupling is restoring the pseudo-Lorenz and internal symmetry. Moreover, it is upgrading the \emph{global} SU(2) internal symmetry to a \emph{local} gauge invariance, provided the defect generated fields are prescribed the proper transformation rules. Of course, the relevance of such symmetry depends on the dynamics of the defect fields not studied here. Therefore the inclusion of the defect dynamics into the consideration is another interesting direction of the development of this work. In particular, it would be interesting to investigate the possibility of having a controlled vacuum expectation value for the scalar sector of the model. This would  be interesting from the theoretical point of view, as well as open new perspectives for practical applications.

\subsection*{Acknowledgments}
I thank  my colleagues and staff from Max-Planck-Insitut f\"{u}r Physik in Munich and Laboratori Nazionali di Frascati, for the warm hospitality and useful discussions. My thanks also extend to respective organizations as well as the Humboldt Foundation for supporting my visit to MPI.  

This work was supported by the Korean NRF research project No.2010-0007637.


\newpage

\appendix
\section{Dirac and Pauli matrices}

\subsubsection*{Reducible Dirac matrices}

We use the following definitions for the Dirac matrices,
\begin{equation}\label{degGamma}
    \gamma^0=\ii \sigma_3 \otimes \sigma_3,\quad 
    \gamma^1=-\sigma_1 \otimes \sigma_3,\quad
    \gamma^2=-\sigma_2 \otimes \I. 
\end{equation}
We use the following definitions of Pauli matrices
\begin{equation}
  \sigma_1=
\begin{pmatrix}
   0 & 1 \\
   1 & 0
\end{pmatrix}, \quad
   \sigma_2=
\begin{pmatrix}
   0 & -\ii \\
  \ii & 0
\end{pmatrix},\quad
  \sigma_3= 
\begin{pmatrix}
  1 & 0 \\
  0 & -1
\end{pmatrix}. 
\end{equation}

The Dirac matrices satisfy the $2+1$-dimensional Clifford algebra,
\begin{equation}
    \{\gamma_\mu,\gamma_\nu\}=2\eta_{\mu \nu}, \qquad \eta_{\mu \nu}=\diag(-1,+1,+1).
\end{equation}

The above Dirac matrices realize a reducible representation. By a unitary transformation we can go to a basis in which the Dirac matrices take the following form,
\begin{equation}\label{propBas}
   \gamma^0 = \ii \sigma_3 \otimes \I,\quad \gamma^i=\sigma_i \otimes \I,
\end{equation}
where $i=1,2$. In what follows, we show that this basis can be obtained by properly labeling the outer products of Pauli matrices multiplied by $-\gamma^0$.

\subsubsection*{SU(2) algebra generators}
The above degeneracy of the Dirac matrices can be described in terms of additional SU(2) symmetry, generated by the following matrices,
\begin{equation}
	T_1=\sigma_2 \otimes \sigma_1,\quad
  T_2=\sigma_2 \otimes \sigma_2,\quad
  T_3=\I \otimes \sigma_3.
\end{equation}
Direct inspection shows that matrices $T_a$, $a=1,2,3$ form an su(2) algebra. They also commute with the Dirac matrices $\gamma^\mu$, 
\begin{equation}
   [T_a, \gamma^\mu]=0,\qquad
   [T_a,T_b]=\ii 2\epsilon_{abc}\tau_c.
\end{equation}

In the diagonal basis of \eqref{propBas} the matrices $T_a$ reduce to  outer products of identity and Puali matrices, $T_a=\I \otimes \tau_a$.\footnote{To avoid the confusion, when speaking about the representation of \eqref{propBas}, we use the notations $\tau_a$ for the Pauli matrices in order to underline their su(2) nature.} 

\subsubsection*{The diagonal basis}
The complete basis for the space of $4 \times 4$ matrices is formed by matrices $\gamma_\mu$, $\gamma_\mu \tau_a$, $\tau_a$ and $1$, where, by slight abuse of notations, we mean irreducible $\gamma$-matrices. We list the basis in the Table \ref{table:DiracBasis}.

\begin{table}[htbp]
	\centering
	\small
	\begin{tabular}{|l|l|l|l|l|}
		\hline
			$\otimes$ & 	$\I$ & 	$\tau_1$ & 	$\tau_2$ & 	$\tau_3$\\
		\hline
			$\I$ & 	$\I \otimes \I$ & 	
	$ \sigma_2\otimes \sigma_1$ & 	
	$\sigma_2\otimes \sigma_2$ & 	
	$\I\otimes \sigma_3$\\
		\hline
			$\gamma_0$ & 
	$\ii\sigma_3 \otimes \sigma_3$	 & 
	$\ii \sigma_1  \otimes \sigma_2$	 & 
	$-\ii \sigma_1 \otimes \sigma_1$ 	 &
	$\ii\sigma_3 \otimes \I$ 	\\
		\hline
			$\gamma_1$ & 
	$-\sigma_1 \otimes \sigma_3$	 & 
	$\sigma_3 \otimes \sigma_2$ 	 &
	$-\sigma_3 \otimes \sigma_1$	 & 
	$-\sigma_1 \otimes\I$	\\
		\hline
			$\gamma_2$ & 
	$-\sigma_2 \otimes \I$ & 
	$-\I \otimes \sigma_1$ & 
	$-\I \otimes \sigma_2$ & 
	$-\sigma_2 \otimes \sigma_3$\\
		\hline
	\end{tabular}
	\caption{The basis for the space of $4 \times 4$ matrices $1=\I \otimes \I$,  $\tau_a$, $\gamma_\mu$, and $\gamma_\mu \tau_a$.}
	\label{table:DiracBasis}
\end{table}

In addition, we need to know the backward `dictionary' to translate an arbitrary product of unity and Pauli matrices multiplied by $-\gamma^0$. These data we summarize in the Table \ref{table:DiracM}.

\begin{table}[htbp]
	\centering
	\small
	\begin{tabular}{|l|l|l|l|l|}
		\hline
			$\otimes$ & 	$\I$ & 	$\sigma_1$ & 	$\sigma_2$ & 	$\sigma_3$\\
		\hline
			$\I$ & 	
	$-\ii \sigma_3\otimes \sigma_3=-\gamma^0$ & 	
	$\sigma_3\otimes \sigma_2=\gamma^1\tau_1$ & 	
	$-\sigma_3\otimes \sigma_1=\gamma^1\tau_2$ & 	
	$-\ii \sigma_3\otimes \I=-\gamma^0\tau_3$\\
		\hline
			$\sigma_1$ & 
	$\sigma_2 \otimes \sigma_3=-\gamma^2 \tau_3$	 & 
	$\ii \sigma_2  \otimes \sigma_2=\ii \tau_2$	 & 
	$-\ii \sigma_2 \otimes \sigma_1=-\ii \tau_1$ 	 & 
	$\sigma_2 \otimes \I=-\gamma^2$ 	\\
		\hline
			$\sigma_2$ & 
	$-\sigma_1 \otimes \sigma_3=\gamma^1$	 & 
	$-\ii \sigma_1 \otimes \sigma_2=-\gamma^0 \tau_1$ 	 & 
	$\ii\sigma_1 \otimes \sigma_1= -\gamma^0 \tau_2$	 & 
	$-\sigma_1 \otimes\I=\gamma^1 \tau_3$	\\
		\hline
			$\sigma_3$ & 
	$-\ii \I \otimes \sigma_3=-\ii \tau_3$ & 
	$\I \otimes \sigma_2=-\gamma^2 \tau_2$ & 
	$-\I \otimes \sigma_1=\gamma^2 \tau_1$ & 
	$-\ii \I \otimes \I=-\ii $\\
		\hline
	\end{tabular}
	\caption{List of outer products of Pauli matrices multiplied by $\gamma_0$, $-\gamma^0\cdot (\sigma_I \otimes \sigma_J)$, $I,J=0,1,2,3$, in terms of gamma matrices and internal symmetry generators. The first factor changes by the row and the second factor by the column.}
	\label{table:DiracM}
\end{table}

\section{Field Tables}\label{appendix:tables}
The low energy coupling to the fermionic field obtained in the main part of the paper has the general form,
\begin{equation}
   \sim \Psi^\dag Z^{IJ} (\sigma_I \otimes \sigma_J) \Psi, \qquad I,J=0,1,2,3.
\end{equation}
Here we are expressing the fields $Z^{IJ}$ in terms of the original defect fields $z_{\mathbf{n}}, z_{\mathbf{n},\hat{\i} }$ and $z_{\mathbf{n}\hat{\i}\hat{\j}}$. 
For the convenience, we are giving these expressions in the form of tables separately for each contribution: local ($\mathrm{n}$), nearest-neighbor ($\mathrm{nn}$) and next-nearest-neighbor ($\mathrm{nnn}$), i.e. such that $Z^{IJ}=Z^{IJ}_{\mathrm{n}}+Z^{IJ}_{\mathrm{nn}}+Z^{IJ}_{\mathrm{nnn}}$, respectively in the tables \ref{table:local}, \ref{table:nearest} and \ref{table:next}. Subscripts ``$+$'' or ``$-$''carried by the fields in the tables below denote that the momentum space modes near the Dirac point $\pm \mathbf{K}$.\footnote{In fact the argument should be $\pm (\mathbf{K}_1-\mathbf{K}_2)=\mp 2\mathbf{K}$, however the periodicity implies that this point is equivalent to $\pm \mathbf{K}$.} No subscript $\pm$ means that the modes near the origin are involved. In the real space this modes are given by the inverse Fourier transform,
\begin{equation}
	z_{*}(\mathbf{r})=\ft{1}{A_{\mathrm{Bz}}}\int\dd^2k z_{*}(\mathbf{k})\e^{-\ii \mathbf{k}\cdot \mathbf{r}} 
\end{equation}
•

The fields appearing in the Table \ref{table:next} are related to the original defect fields (clarify  \eqref{z-fields}) by $\mathbf{z}=\ft{1}{2\pi} \sum_i z_{\hat{l}_i } \hat{k}_i$, where $\hat{k}_i$  are vectors of the  dual basis. In coordinates, the vector $\mathbf{z}$ is represented by,
\begin{equation}
	z_{x}=(z_{\hat{2}} +z_{\hat{3}}), \qquad
	z_{y}=-\sqrt{3} (z_{\hat{2}} -z_{\hat{3}}).
\end{equation}
In addition, we use the following notations for the fields appearing in the tables,
\begin{align}
	z^0_{x}&=\ft{1}{2}(z_{A \hat{1}\hat{2}}+z_{B \hat{1}\hat{2} } 
	+z_{A \hat{1} \hat{3} }+z_{B \hat{1} \hat{3} }),\\
	 z^0_{y}&=-\ft{\sqrt{3}}{2}(z_{A \hat{1}\hat{2}}+z_{B \hat{1}\hat{2} } 
	-z_{A \hat{1} \hat{3} }-z_{B \hat{1} \hat{3} }),\\
	z^3_{x}&=\ft{1}{2}(z_{A \hat{1}\hat{2}}-z_{B \hat{1}\hat{2} } 
	+z_{A \hat{1} \hat{3} }-z_{B \hat{1} \hat{3} }),\\
	z^3_{y}&=-\ft{\sqrt{3}}{2}(z_{A \hat{1}\hat{2}}-z_{B \hat{1}\hat{2} } 
	-z_{A \hat{1} \hat{3} }+z_{B \hat{1} \hat{3} }),
\intertext{as well as}
	z^{\prime 0}&=\ft{1}{2} (z_{A \hat{2}\hat{3} }+z_{B \hat{2}\hat{3} }),\\
	z^{\prime 3}&=\ft{1}{2} (z_{A \hat{2}\hat{3} }-z_{B \hat{2}\hat{3} }).
\end{align}


\begin{table}[htbp]
	\centering
	\normalsize
	\begin{tabular}{|l|l|l|l|l|}
		\hline
			$\otimes$ & 	$\I$ & 	$\sigma_1$ & 	$\sigma_2$ & 	$\sigma_3$\\
		\hline
			$\I$ & $\ft{1}{2}(z_A+z_B) $	 & $\ft{1}{4}(z_{A+}+z_{A-}+z_{B+}+z_{B-}) $	 &$ \ft{\ii}{4}(z_{A-} -z_{A+}+z_{B-}-z_{B+})$ 	 & $0$	\\
		\hline
			$\sigma_1$ & $\ft{1}{2}(z_{\hat{1}}+\bar{z}_{\hat{1}})$  	 & $\ft{1}{4}(z_{\hat{1}-}+\bar{z}_{\hat{1}-}+z_{\hat{1}+}+\bar{z}_{\hat{1}+})$	 & $\ft{\ii}{4}(z_{\hat{1}-}+\bar{z}_{\hat{1}-}-z_{\hat{1}+}-\bar{z}_{\hat{1}+})$ 	 & $0$	\\
		\hline
			$\sigma_2$ & $\ft{\ii}{2}(z_{\hat{1}}-\bar{z}_{\hat{1}}) $	 & $\ft{\ii}{4}(z_{\hat{1}-}-\bar{z}_{\hat{1}-}+z_{\hat{1}+}-\bar{z}_{\hat{1}+}) $	 & $-\ft{1}{4}(z_{\hat{1}-}-\bar{z}_{\hat{1}-}-z_{\hat{1}+ }+\bar{z}_{\hat{1}+ } ) $	 & $0$ 	\\
		\hline
			$\sigma_3$ & $\ft{1}{2}(z_A-z_B) $	 & $\ft{1}{4}(z_{A-}-z_{B-}+z_{A+}-z_{B+})$ 	 & $\ft{\ii}{4}(z_{A-}-z_{B-}-z_{A+}+z_{B+}) $ 	 & $0$	\\
		\hline
	\end{tabular}
	\caption{The local contribution $Z_{\mathrm{n}}^{IJ}$. The first index is counting rows while the second is counting columns. The fields are listed according to the definition \eqref{z-fields}}
	\label{table:local}
\end{table}


\begin{table}[htbp]
	\centering
	\footnotesize
	\begin{tabular}{|l|l|l|l|l|}
		\hline
			$\otimes$ & 	$\I$ & 	$\sigma_1$ & 	$\sigma_2$ & 	$\sigma_3$\\
		\hline
			$\I$ & $0$	 & $0$	 & $0$	 & $0$	\\
		\hline
			$\sigma_1$ & $-\ft{3}{8} (z_x+\bar{z}_x) $	 & 
$\begin{array}{c}
	-\ft{3}{8}[z_{x-}+\bar{z}_{x-}+z_{x+}+\bar{z}_{x+}\\
	+\ii (z_{y+}+\bar{z}_{y+} -z_{y-}-\bar{z}_{y-})]
\end{array}$	 & 	
$\begin{array}{c}
	-\ft{3\ii}{8}[z_{x-}+\bar{z}_{x-}-z_{x+}-\bar{z}_{x+}\\
	-\ii (z_{y+}+\bar{z}_{y+} +z_{y-}+\bar{z}_{y-})]
\end{array}$
 & $-\ft{3\ii}{8}(z_{y}-\bar{z}_{y})$ 	\\
		\hline
			$\sigma_2$ & $-\ft{3\ii}{8} (z_x-\bar{z}_x) $	 & 	
$\begin{array}{c}
	-\ft{3\ii}{8}[z_{x-}-\bar{z}_{x-}+z_{x+}-\bar{z}_{x+}\\
	+\ii (z_{y+}-\bar{z}_{y+} -z_{y-}+\bar{z}_{y-})]
\end{array}$
 & 
$\begin{array}{c}
	\ft{3}{8}[z_{x-}-\bar{z}_{x-}-z_{x+}+\bar{z}_{x+}\\
	-\ii (z_{y+}-\bar{z}_{y+}+z_{y-}-\bar{z}_{y-})]
\end{array}$
	 &  $\ft{3}{8}(z_{y}+\bar{z}_{y}) $	\\
		\hline
			$\sigma_3$ & $0$	 & $0$	 & $0$	 & $0$	\\
		\hline
	\end{tabular}
	\caption{The local contribution $Z_{\mathrm{nn}}^{IJ}$. The first index is counting rows while the second is counting columns.}
	\label{table:nearest}
\end{table}


\begin{table}[htbp]
	\centering
	\scriptsize
	\begin{tabular}{|l|l|l|l|l|}
		\hline
			$\otimes$ & 	$\I$ & 	$\sigma_1$ & 	$\sigma_2$ & 	$\sigma_3$\\
		\hline
			$\I$ & 
$ 
\begin{array}{c}
	-\ft{3}{4}(z^0_{x}+\bar{z}^0_x)\\
	-\ft{1}{4}(z^{\prime 0}+\bar{z}^{\prime 0})  
\end{array}
$	 & 
$
\begin{array}{c}
	-\ft{3}{4}[z_{x-}^0+\bar{z}_{x-}^0+z_{x+}^0+\bar{z}_{x+}^0\\
	+\ii(z_{y+}^0+\bar{z}_{y+}^0-z_{y-}^0-\bar{z}_{y-}^0)]\\
	+\ft{1}{4}(z_{-}^{\prime 0} +z_{+}^{\prime 0}+\bar{z}_{-}^{\prime 0}+\bar{z}_{+}^{\prime 0})
\end{array}
$	 & 
$
\begin{array}{c}
	-\ft{3\ii}{4}[z_{x-}^0+\bar{z}_{x-}^0-z_{x+}^0-\bar{z}_{x+}^0\\
	-\ii(z_{y+}^0+\bar{z}_{y+}^0+z_{y-}^0+\bar{z}_{y-}^0)]\\
	+\ft{\ii}{4}(z_{-}^{\prime 0} +z_{+}^{\prime 0}-\bar{z}_{-}^{\prime 0}-\bar{z}_{+}^{\prime 0})
\end{array}
$
	 &  
$
\begin{array}{c}
	-\ft{3\ii}{4}(z_{y}^0-\bar{z}_{y}^0)\\
	+\ft{\ii \sqrt{3}}{4}(z^{\prime 0}-\bar{z}^{\prime 0})  
\end{array}
$
	\\
		\hline
			$\sigma_1$ & $0$	 & $0$	 & $0$	 & $0$	\\
		\hline
			$\sigma_2$ & $0$	 & $0$	 & $0$	 & $0$	\\
		\hline
			$\sigma_3$ & 	
$
\begin{array}{c}
-\ft{3}{4}(z^3_{x}+\bar{z}^3_{x})\\
-\ft{1}{4}(z^{\prime 3}+\bar{z}^{\prime 3})  
\end{array}
$
 &
$
\begin{array}{c}
	-\ft{3}{4}[z_{x-}^3+\bar{z}_{x-}^3+z_{x+}^3+\bar{z}_{x+}^3\\
	+\ii(z_{y+}^3+\bar{z}_{y+}^3-z_{y-}^3-\bar{z}_{y-}^3)]\\
	+\ft{1}{4}(z_{-}^{\prime 3} +z_{+}^{\prime 3}+\bar{z}_{-}^{\prime 3}+\bar{z}_{+}^{\prime 3})
\end{array}
$
 	 &
$
\begin{array}{c}
	-\ft{3\ii}{4}[z_{x-}^3+\bar{z}_{x-}^3-z_{x+}^3-\bar{z}_{x+}^3\\
	-\ii(z_{y+}^3+\bar{z}_{y+}^3+z_{y-}^3+\bar{z}_{y-}^3)]\\
	+\ft{\ii}{4}(z_{-}^{\prime 3} +z_{+}^{\prime 3}-\bar{z}_{-}^{\prime 3}-\bar{z}_{+}^{\prime 3})
\end{array}
$
 	 &
$
\begin{array}{c}
	-\ft{3\ii}{4}(z_{y}^3-\bar{z}_{y}^3)\\
	+\ft{\ii \sqrt{3}}{4}(z^{\prime 3}-\bar{z}^{\prime 3})  
\end{array}
$ 	\\
		\hline
	\end{tabular}
	\caption{The local contribution $Z_{\mathrm{nnn}}^{IJ}$. The first index is counting rows while the second is counting columns.}
	\label{table:next}
\end{table}

\bibliographystyle{unsrt}
\bibliography{graphene_phonon}
	
\end{document}